\documentstyle[prl,aps,amsfonts,amssymb,floats,epsfig]{revtex}

\begin{document}
\draft
\twocolumn[\hsize\textwidth\columnwidth\hsize\csname
@twocolumnfalse\endcsname
\title{
Magnetic excitations in two-leg spin 1/2 ladders:
experiment and theory
}
\author{
M. Gr\"{u}ninger,$^{1}$ M. Windt,$^{1}$ T. Nunner,$^{2}$ C.
Knetter,$^{3}$ K.P. Schmidt,$^{3}$ G.S.
Uhrig,$^{3}$ T. Kopp,$^{2}$ A. Freimuth,$^{1}$
\\
U. Ammerahl,$^{1,5}$ B. B\"{u}chner,$^{1,4}$ and A. Revcolevschi$^{5}$}
\address{
$^1$\it II. Physikalisches Institut, Universit\"{a}t zu K\"{o}ln,
50937 K\"{o}ln, Germany
\\
{\rm $^2\!$} Experimentalphysik VI, Universit\"{a}t Augsburg,
86135 Augsburg, Germany
\\
{\rm $^3\!$} Institut f\"{u}r Theoretische Physik, Universit\"{a}t zu K\"{o}ln,
50937 K\"{o}ln, Germany
\\
{\rm $^4\!$} II. Physikalisches Institut, RWTH-Aachen, 52056 Aachen, Germany
\\
{\rm $^5\!$} Laboratoire de Chimie des Solides,
Universit\'e Paris-Sud, 91405 Orsay C\'edex, France
}
\date{September 26, 2001}
\maketitle
\begin{abstract}
Magnetic excitations in two-leg $S$=1/2 ladders are studied both
experimentally and theoretically. Experimentally, we report
on the reflectivity, the transmission and the
optical conductivity $\sigma(\omega)$ of undoped
La$_x$Ca$_{14-x}$Cu$_{24}$O$_{41}$ for $x$=4, 5, and 5.2. Using
two different theoretical approaches (Jordan-Wigner fermions and
perturbation theory), we calculate the dispersion of the
elementary triplets, the optical conductivity and the
momentum-resolved spectral density of two-triplet excitations for
$0.2 \leq J_\parallel/J_\perp \leq 1.2$. We discuss
phonon-assisted two-triplet absorption, the existence of
two-triplet bound states, the two-triplet continuum, and the size
of the exchange parameters.
\end{abstract}
\vskip2pc]
\narrowtext

{\bf I. Introduction}

\bigskip

Low-dimensional quantum antiferromagnets offer a diverse view on
quantum fluctuations at work. Particular interest has focused on
two-leg spin 1/2 ladders, which show a spin liquid ground state
with a spin gap to the lowest excited state, and
superconductivity under pressure upon hole doping
\cite{uehara}. In undoped two-leg ladders the existence of
two-``magnon'' (two-triplet) bound states was predicted theoretically
by a number of groups
\cite{uhrigschulz,damle,sushkovPRL,jurecka,monienPRL,zheng}.
Recently, we have reported on the observation of a two-triplet bound
state with $S_{\rm tot}$=0 in the optical conductivity spectrum of
(La,Ca)$_{14}$Cu$_{24}$O$_{41}$ \cite{windt}, a compound with
stacked layers of Cu$_2$O$_3$ ladders and of CuO$_2$ chains.
Concerning the peak positions and the line shape of the bound
states, good agreement was achieved between experiment and two
different theoretical approaches, namely, Jordan-Wigner fermions
and perturbation expansion using unitary transformations
\cite{windt}. Using the second approach, some of us have
discussed the momentum-resolved spectral densities and the
relationship between fractional (spinons) and integer excitations
(triplets) in two-leg spin 1/2 ladders \cite{knetterPRL}. In
these proceedings, we discuss recent experimental and theoretical
progress that we have achieved.
\bigskip

{\bf II. Experiment}

\bigskip

Single crystals of La$_x$Ca$_{14-x}$Cu$_{24}$O$_{41}$ were grown
by the travelling solvent floating zone method \cite{udo}. These
so called telephone number compounds have attracted particular
interest due to the possibility of hole doping. Here, we are
interested in the magnetic properties of nominally undoped
samples, i.e. Cu$^{2+}$, which corresponds to $x$=6. Single phase
crystals could only be synthesized for $x \leq 5.2$ \cite{udo}.
The samples studied here with $x$=5.2, 5, and 4 on average
contain $n$=0.8/24, 1/24 and 2/24 holes per Cu, respectively.
However, x-ray absorption data show that for these low doping
levels the holes are located within the chains \cite{nuecker}.
Thus, we consider the ladders to be undoped \cite{windt}.

In order to determine the optical conductivity $\sigma (\omega)$
we have measured both transmission and reflection data between
500 and $12\,000$ cm$^{-1}$ on a Fourier spectrometer. The
transmittance $T(\omega)$ and the reflectivity
$R(\omega)$ were calibrated against a reference aperture and a
gold mirror, respectively. The optical
conductivity $\sigma(\omega)=n\kappa
\omega /2\pi$ was determined by inverting \cite{diplom,choi}
\begin{equation}
R(\omega)=\left[(n-1)^2+\kappa^2 \right] \, / \,
\left[ (n+1)^2+\kappa^2 \right]
\; ,
\end{equation}
\begin{equation}
T(\omega)=\left[ (1-R)^2\Phi \right] \, / \, \left[ 1-(R\Phi)^2
\right]
\; ,
\end{equation}
\begin{equation}
\Phi(\omega)=\exp(-2\omega \kappa d /c)=\exp(-\alpha d ) \; ,
\end{equation}
where $n$ denotes the index of refraction, $\kappa$ the extinction
coefficient, $\alpha$ the absorption coefficient, $c$ the velocity
of light and $d$ the thickness of the transmission sample [$R(\omega)$
denotes the single bounce reflectivity and hence needs to be measured
on a thick (``semi-infinite''), opaque sample].
Equation 2 is obtained for a sample with parallel surfaces by adding
up the intensities of all
multiply reflected beams {\em incoherently}, i.e., by neglecting
interference effects. Experimentally, this condition is realized
either if the sample surfaces are not perfectly parallel or by
smoothing out the Fabry-Perot interference fringes by means of
Fourier filtering. In case of weak absorption $\kappa \ll n$, the
reflectivity is entirely determined by $n$ and not suitable to
derive $\kappa$ by using a Kramers-Kronig transformation.
At the same time, $\kappa$ can be determined very accurately from the
transmission. Since $T(\omega)$ depends exponentially on $\kappa \cdot d$,
the appropriate choice of $d$ is essential. Furthermore, the thickness $d$
determines the effective ``resolution'' given by the interference fringes.
The period of the fringes is given by $\Delta \tilde{\nu}=(2nd)^{-1}$, where
$\Delta\tilde{\nu}$ is given in wave numbers if $d$ is given in units of cm.
Thus, a large value of $d$ has the advantage of narrow fringes with a
small amplitude, but strongly restricts the maximum value of
$\kappa$ that can be observed.

\begin{figure}[t]
\centerline{\psfig{figure=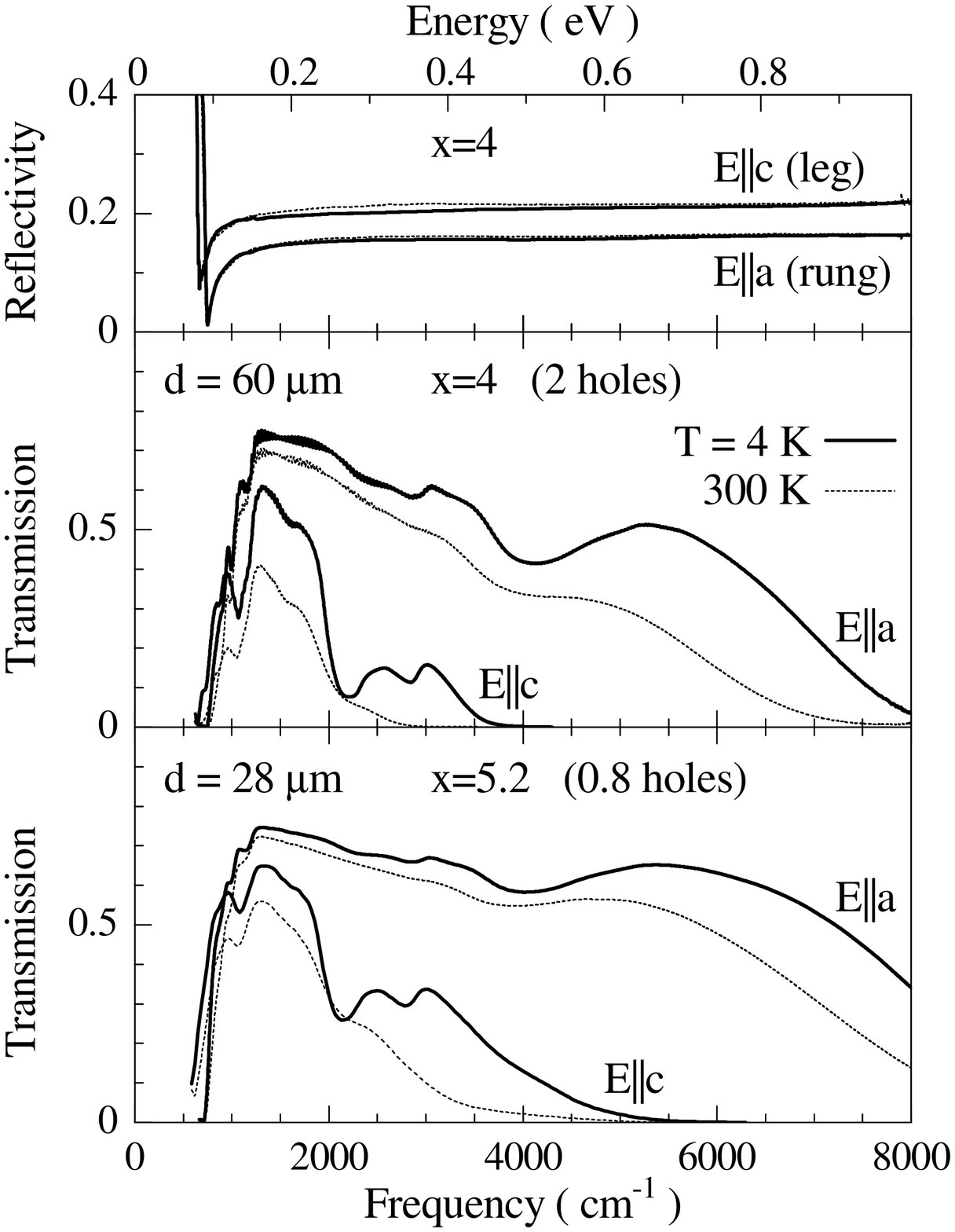,width=7.2cm,clip=}}
  \caption{Midinfrared reflectivity and transmission of
           La$_x$Ca$_{14-x}$Cu$_{24}$O$_{41}$ at 4 and 300 K.\@}
\label{rt}
\end{figure}

The top panel of Fig.\ \ref{rt} shows the reflectivity measured on a
0.8~mm thick sample ($x$=4) for polarization of
the electrical field both parallel to the rungs and to the legs.
At about 600-700~cm$^{-1}$ one can observe the oxygen bond
stretching phonon mode. At higher frequencies, the reflectivity is
featureless, which is characteristic for the weak absorption
regime below the gap of an insulator. We plotted the data for
both 4~K (thick lines) and 300~K (thin dashed lines), but in
$R(\omega)$ they are almost indistinguishable. The
different absolute values of the two polarization directions
reflect the difference in $n$, namely, $n_a \approx 2.3$ and
$n_c \approx 2.6$. The weak absorption features we are looking
for can only be detected in a transmission experiment, which also
reveals a strong temperature dependence. The middle and bottom
panels of Fig.\ \ref{rt} show $T(\omega)$ measured on thin single
crystals with $x$=4 ($d$=60$\mu$m) and $x$=5.2 ($d$=28$\mu$m).
The data in the middle panel were
measured with a resolution of 4~cm$^{-1}$ and show
only small interference fringes, most probably because the sample
surfaces were not perfectly parallel. The sample with $x$=5.2
(bottom panel) showed strong interference fringes with a period of
$\approx 70$~cm$^{-1}$, which have been removed by Fourier filtering.

\begin{figure}[t]
\centerline{\psfig{figure=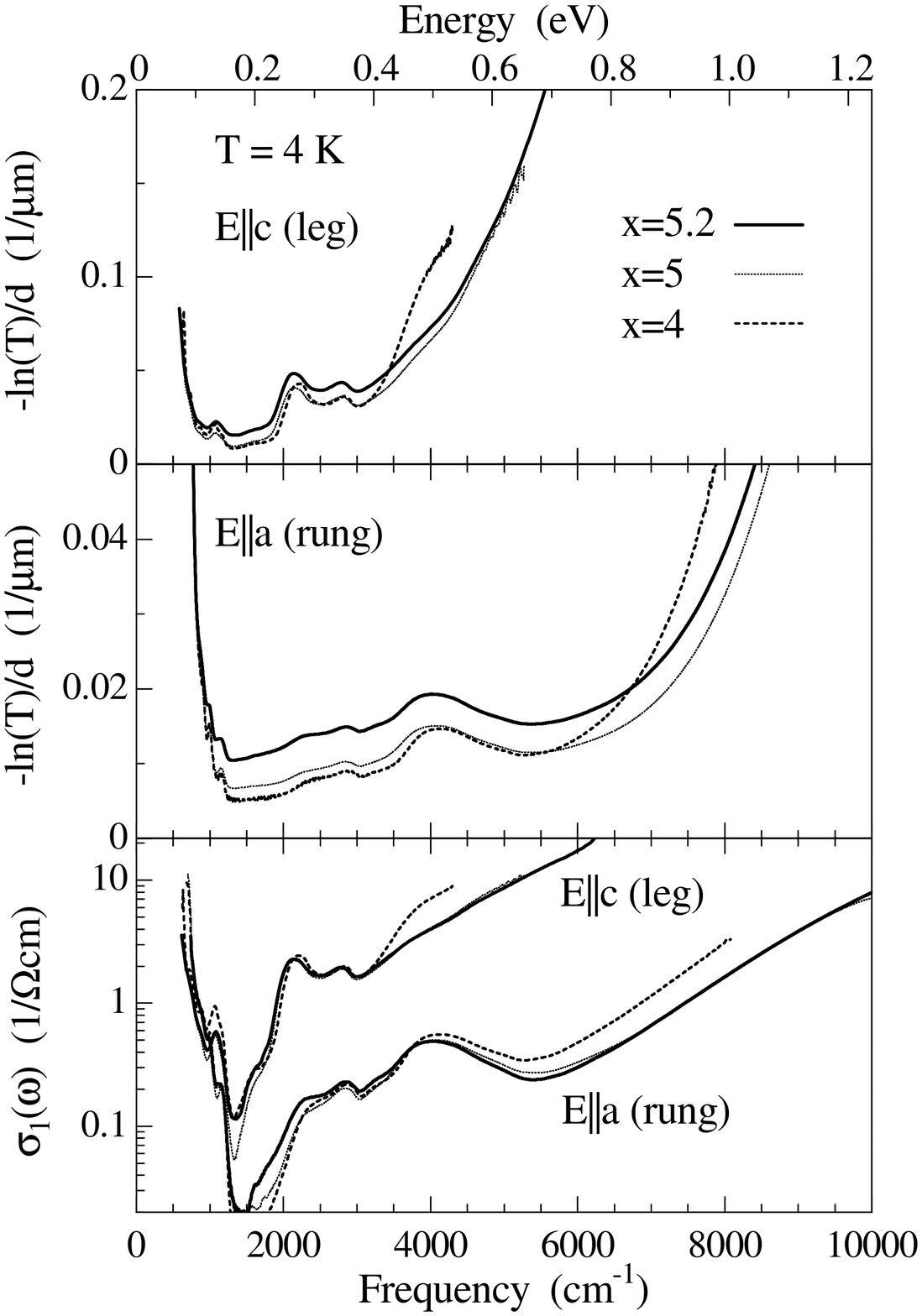,width=7.2cm,clip=}}
  \caption{Bottom panel: $\sigma_1(\omega)$ of La$_x$Ca$_{14-x}$Cu$_{24}$O$_{41}$
  ($x$=4, 5, and 5.2) at 4 K, calculated from transmission and reflectivity after
  Eq.\ 2.
  Top and middle panel: Plotting the logarithm of the transmission divided by the
  sample thickness already gives a qualitative estimate of the absorption features
  without knowledge of the reflectivity.
  }
\label{lnt}
\end{figure}

In order to obtain a qualitative estimate of the absorption features it is common to approximate
Eq.\ 2 by
\begin{equation}
T(\omega) \approx (1-R)^2 \Phi \; .
\end{equation}
In this case one obtains the absorption coefficient $\alpha(\omega)$ as
\begin{equation}
\alpha(\omega)\approx - \ln(T) / d \; + \; 2 \ln(1-R) / d \; .
\end{equation}
Neglecting the almost constant second term, we plot $-\ln(T(\omega))/d$ for
$x$=4, 5, and 5.2 at 4~K in the top ($E\parallel c$) and middle panels
($E\parallel a$) of Fig.\ \ref{lnt}. For comparison, the optical conductivity
$\sigma(\omega)$ as derived by inverting Eq.\ 2 without further assumptions is
displayed in the bottom panel of Fig.\ \ref{lnt}. In $\sigma(\omega)$ the
curves for $x$=5 and $x$=5.2 almost fall on top of each other. Note that the
erroneous discrepancy between these two data sets in $-\ln(T)/d$ is {\em not}
due to a difference in $R(\omega)$, but due to the approximation used in Eq.\ 4.
A precise and reliable determination of $\sigma(\omega)$ in the case of small
absorption thus requires the measurement of {\em both} the transmission $T$
and the reflectivity $R$ and the use of Eq.\ 2.

\bigskip

\begin{figure}[t]
\centerline{\psfig{figure=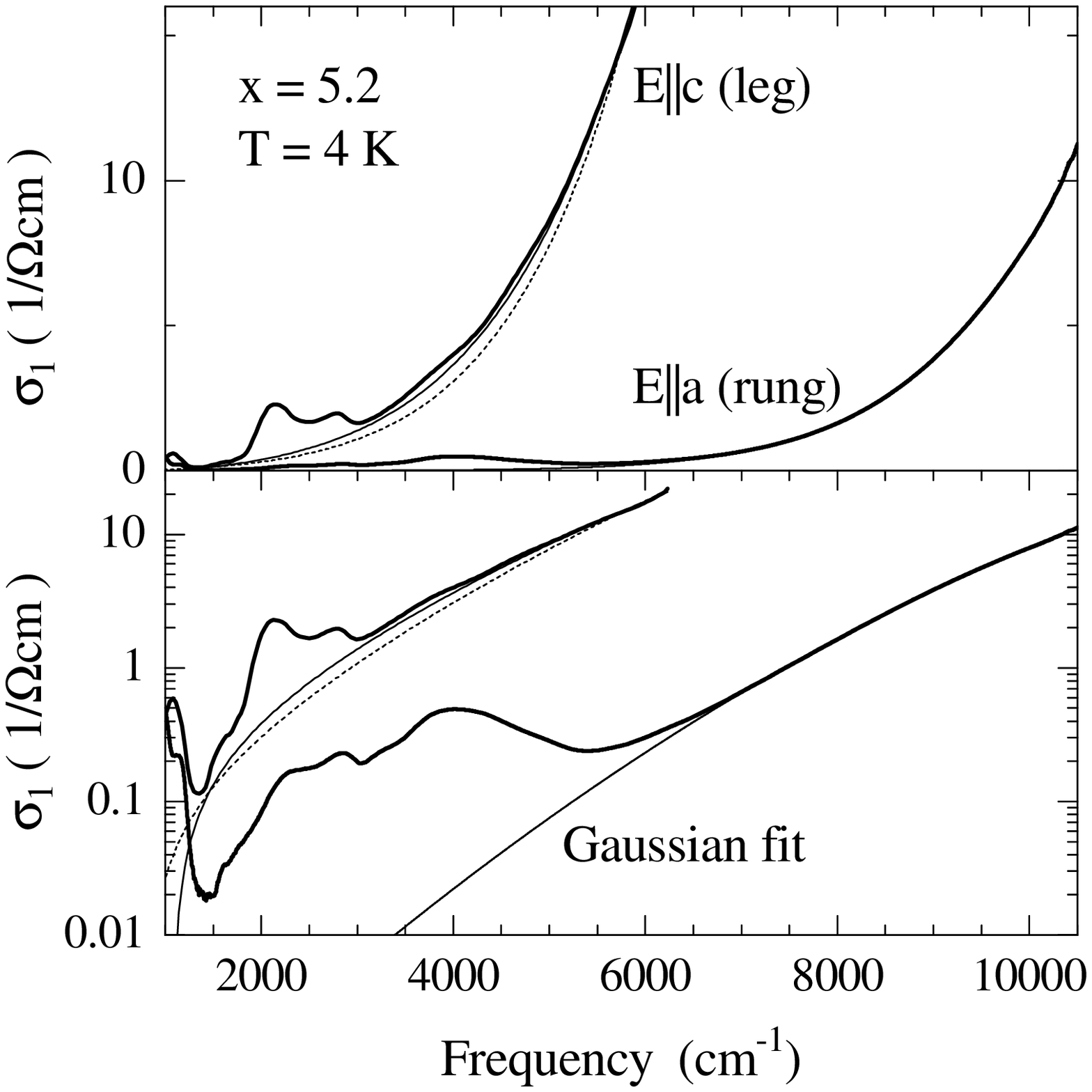,width=7.2cm,clip=}}
  \caption{Optical conductivity $\sigma_1(\omega)$ of La$_{5.2}$Ca$_{8.8}$Cu$_{24}$O$_{41}$ at 4 K
  (thick lines) on a linear (top panel) and on a logarithmic scale (bottom panel).
  The thin lines show Gaussian fits of the electronic background.
  For $E\parallel a$ the fit is unambiguous.
  For $E\parallel c$ two different background fits are shown in order to illustrate
  the uncertainty (see Fig.\ \ref{exptheo}).
  }
\label{lnsigma}
\end{figure}

{\bf III. Magnetic contribution to $\sigma(\omega)$ }

\bigskip

The spectra of $\sigma(\omega)$ can be divided into three
different regimes. The absorption below $\approx$~1300~cm$^{-1}$
can be attributed to phonons and multi-phonon bands. The strong
increase at high frequencies is due to an electronic background,
which most probably has to be identified with the onset of
charge-transfer excitations. We focus on the features in the
intermediate frequency range. In order to analyze them we used
Gaussian fits \cite{expfit} to subtract the electronic background
(thin lines in Fig.\ \ref{lnsigma}). For $E\!\parallel
\! a$, the background can be determined unambiguously by fitting
the measured data for $\omega \! > \! 7000$ cm$^{-1}$. After
subtraction of the background the $\sigma_a (\omega)$ curves are
nearly independent of $x$, which corroborates the assumption that
the ladders are undoped. For $E\!\parallel\! c$, the higher
absorption complicates the determination of the background
considerably. We were able to measure the transmission up to
$\approx 6200$ cm$^{-1}$ (in case of the $x$=5.2 sample with
$d$=28~$\mu$m). Although this is a significant
improvement compared to our earlier work \cite{windt} which was
restricted to $\omega \lesssim 5000$ cm$^{-1}$ ($x$=5), it is not
yet sufficient for an unambiguous determination of the background.
Therefore we show two different possibilities for Gaussian fits
of the $c$-axis background in Fig.\ \ref{lnsigma}.

\begin{figure}[t]
\centerline{\psfig{figure=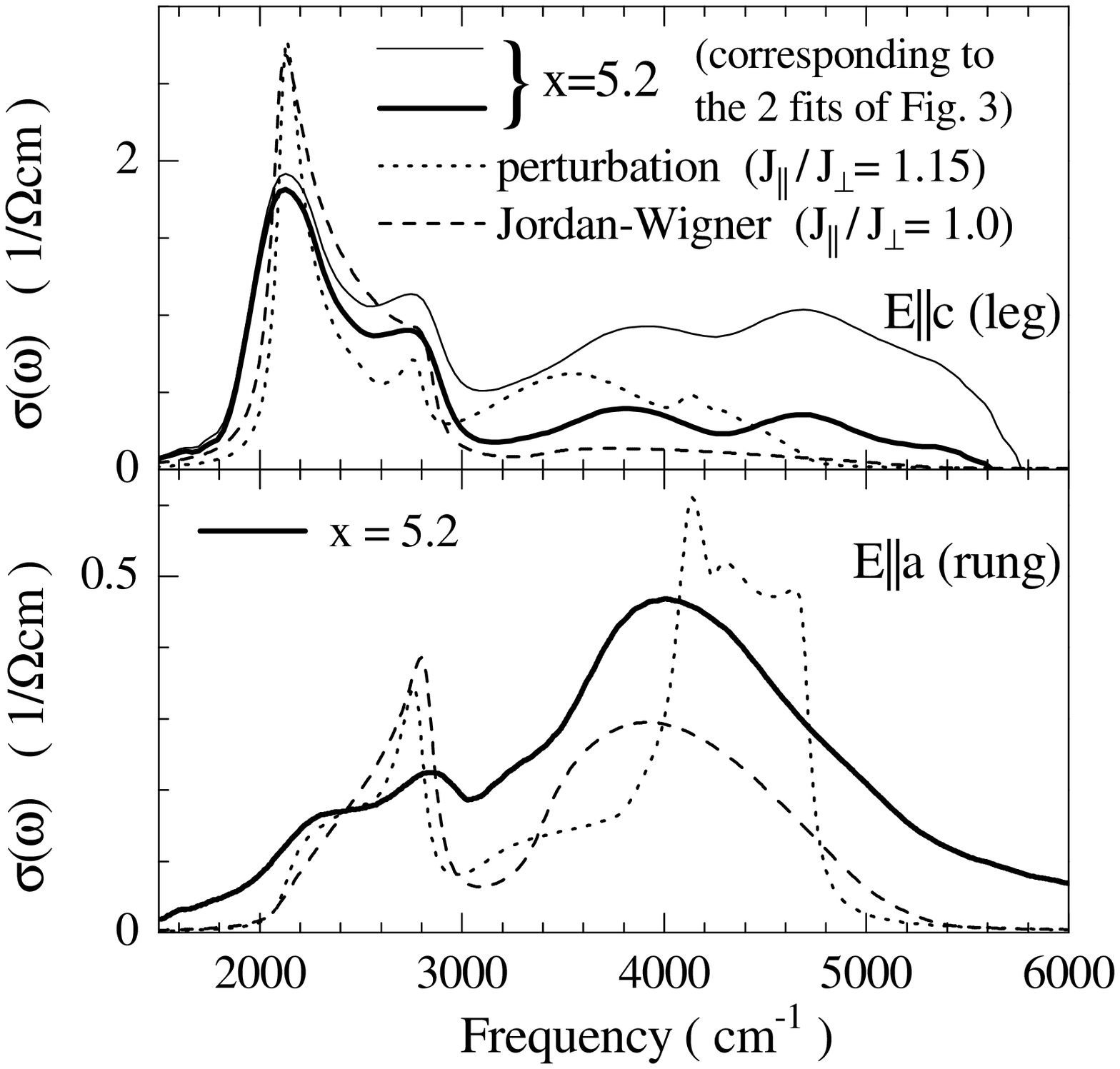,width=7.2cm,clip=}}
  \caption{Comparison of the magnetic contribution to the optical conductivity of
  La$_{5.2}$Ca$_{8.8}$Cu$_{24}$O$_{41}$ at 4 K (solid lines) and of the calculated
  results obtained with Jordan-Wigner fermions ($J_\parallel/J_\perp \! = \! 1.0$;
  $J_\parallel \! = \! 1100$~cm$^{-1}$; dashed)
  and with optimized perturbation theory ($J_\parallel/J_\perp \! = \! 1.15$;
  $J_\parallel \! = \! 1080$~cm$^{-1}$; dotted). In the top panel, the two different
  experimental curves (thin and thick solid lines) correspond to the two alternative
  background fits of Fig.\ \ref{lnsigma}.
}
\label{exptheo}
\end{figure}

The estimates of the magnetic contribution to $\sigma(\omega)$ which we have
obtained this way are plotted in Fig.\ \ref{exptheo} (thin and thick solid
lines). In Ref.\ \onlinecite{windt} we interpreted these features in terms of
phonon-assisted two-triplet absorption \cite{lorenzana95}, which was used to
describe $\sigma_1$ of the undoped 2D cuprates (e.g.\ $\rm YBa_2Cu_3O_6$
\cite{gruen}) and of the 1D S=1/2 chain $\rm Sr_2CuO_3$ \cite{lorenzana97}.
Because of spin conservation two triplets are excited. The simultaneous
excitation of a phonon provides the symmetry breaking necessary to bypass the
selection rule and at the same time takes care of momentum conservation
\cite{windt,lorenzana95,gruen}. This process is the lowest order magnetic
response possible in infrared absorption. The two peaks between 2000 and
3000~cm$^{-1}$ (see Fig.\ \ref{exptheo}) can be identified with the 1D
van Hove singularities in the density of states of the strongly dispersing
two-triplet bound state with $S_{\rm tot}$=0 \cite{windt} (see below).
The absorption at higher energies is attributed to the two-triplet continuum.
The improved estimate of the background as compared to Ref.\ \onlinecite{windt}
allows us to detect {\em two} peaks within the continuum at about 3800 and
4700~cm$^{-1}$ for $E\parallel c$. The precise determination of the spectral
weight of these high energy features requires additional work. Note that also
the $a$-axis spectrum displays a shoulder at high energies
($\approx 5000$~cm$^{-1}$), although it is quite weak.
\bigskip

{\bf IV. Theory}
\bigskip

\begin{figure}[t]
\centerline{\psfig{figure=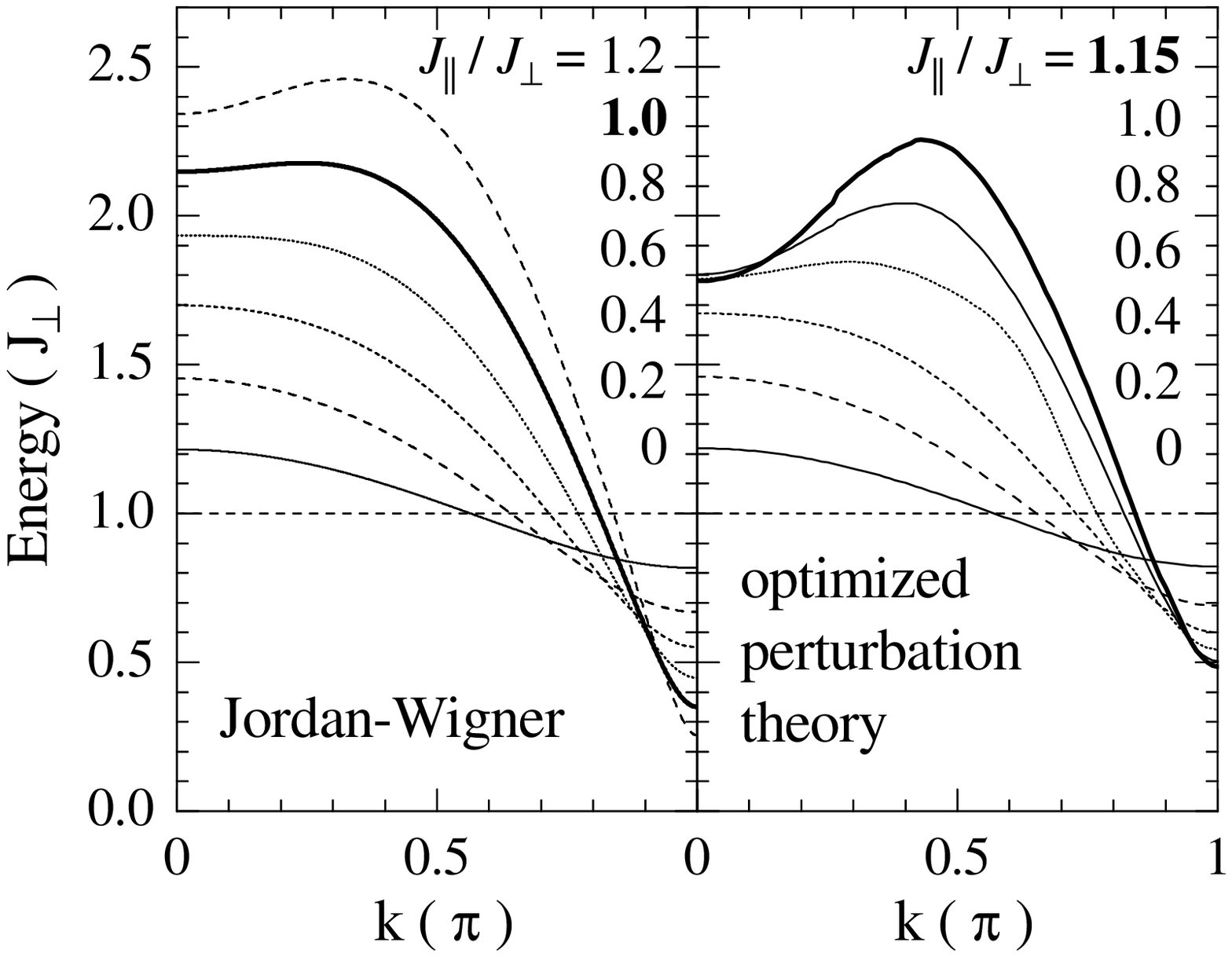,width=7.2cm,clip=}}
  \caption{Elementary triplet dispersion for $0\leq J_\parallel / J_\perp \leq 1.2$.
  For each theory, the thick lines denote the coupling ratio that offers the best
  description of the optical conductivity of (La,Ca)$_{14}$Cu$_{24}$O$_{41}$
  (see Fig.\ \ref{exptheo}).
}
\label{disp}
\end{figure}

Antiferromagnetic $S=1/2$ two-leg Heisenberg ladders are represented
by the Hamiltonian
\begin{equation}
{\cal H}=\sum_{i} \left\{
J_\parallel ({\bf S}_{1,i}{\bf S}_{1,i+1} + {\bf S}_{2,i}{\bf S}_{2,i+1})
+ J_\perp {\bf S}_{1,i}{\bf S}_{2,i} \right\}
,
\label{hamilton}
\end{equation}
where $J_\perp$ and $J_\parallel$ denote the rung and leg
couplings, respectively. For $J_\parallel$=0 one can excite local
rung singlets to rung triplets which become dispersive on finite
$J_\parallel$ (see Fig.\ \ref{disp}). For $J_\perp$=0 the $S$=1
chain excitations decay into asymptotically free $S$=1/2 spinons.
An  intuitive picture of  the ``magnons'' (elementary triplets)
for $J_\perp$, $J_\parallel \neq$ 0 can be obtained from both
limits: the elementary excitations are either dressed triplet
excitations or pairs of bound spinons with a finite gap $\Delta$
as long as $J_\perp \! > \! 0$ \cite{knetterPRL}.

Here, we present two theoretical approaches to describe the excitations of
the ladder. One approach makes use of the Jordan-Wigner transformation to
rewrite the spins as fermions with a long-ranged phase factor. Expanding the
phase factor yields new interaction terms between the fermions. The resulting
interacting fermion problem is treated diagrammatically. A similar treatment
works very well for a 1D chain. The other approach uses an extrapolated
perturbation in $J_\parallel/J_\perp$ and separates contributions of different
triplet number by continuous unitary transformations
\cite{KnetterEuro,knetterPRL}. Both methods are controlled in the sense that
they become exact on $J_\parallel /J_\perp \rightarrow 0$. In fact, comparing
the results of these two methods for the dispersion of the elementary triplets
(see Fig.\ \ref{disp}), the agreement is excellent for
$J_\parallel / J_\perp \leq 0.6$.

\begin{figure}[t]
\centerline{\psfig{figure=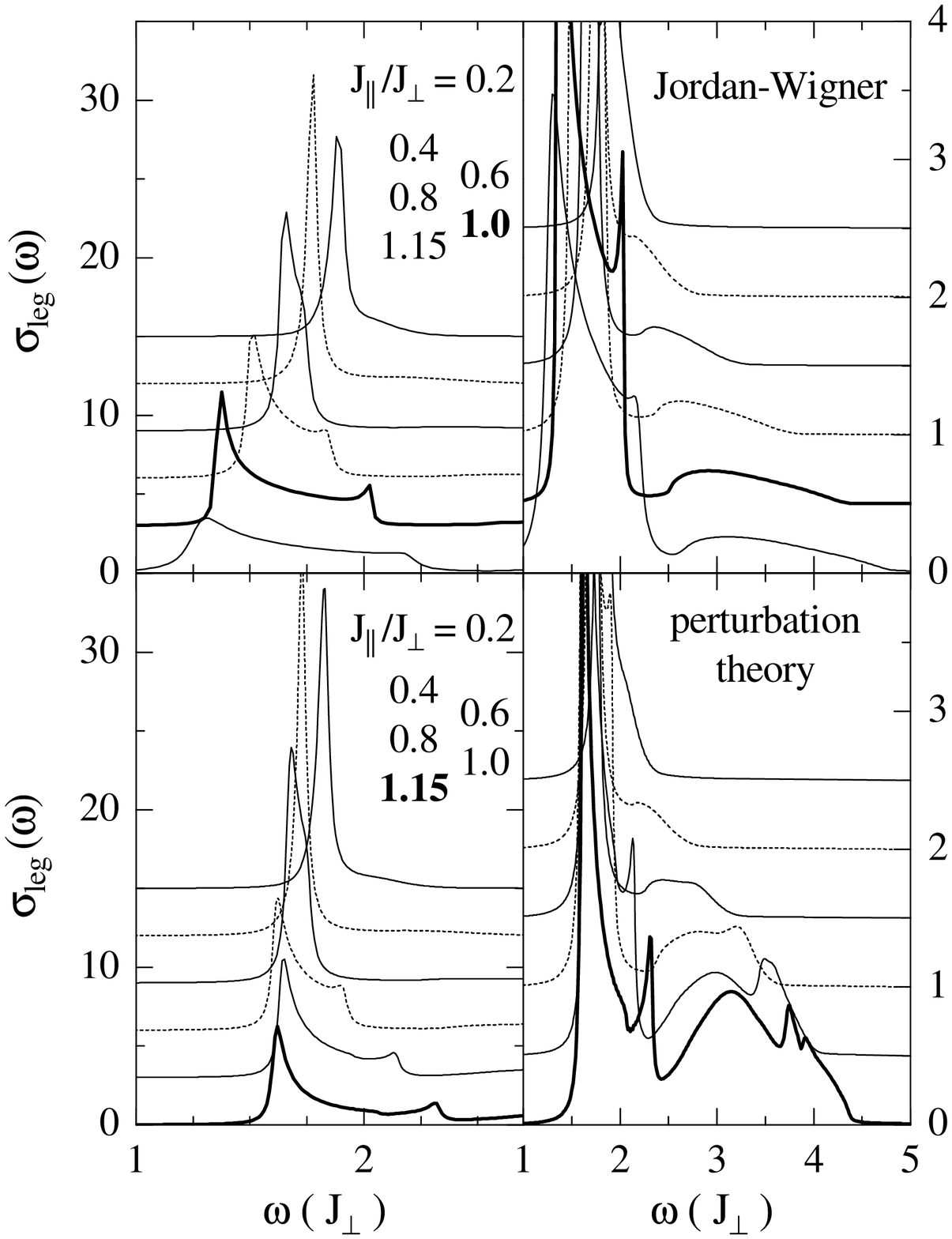,width=7.2cm,clip=}}
  \caption{Comparison of the optical conductivity for $E\parallel c$ (leg) calculated
  with Jordan-Wigner fermions (top panels) and with optimized perturbation theory (bottom)
  for $0\leq J_\parallel / J_\perp \leq 1.15$.
  The left panels focus on the bound state, whereas the right panels emphasize the
  two-triplet continuum on an enlarged $y$-scale. For clarity, the curves were shifted
  with respect to each other by 3 (0.5) $(\Omega$cm$)^{-1}$ in the left (right) panels.
  For each theory, the thick lines denote the coupling ratio that offers the best
  description of the bound states in $\sigma(\omega)$ of
  (La,Ca)$_{14}$Cu$_{24}$O$_{41}$ (see Fig.\ \ref{exptheo}). For comparison with experiment,
  the phonon frequency still has to be added (here: 600~cm$^{-1}$).
}
\label{pint}
\end{figure}

\begin{figure}[t]
\centerline{\psfig{figure=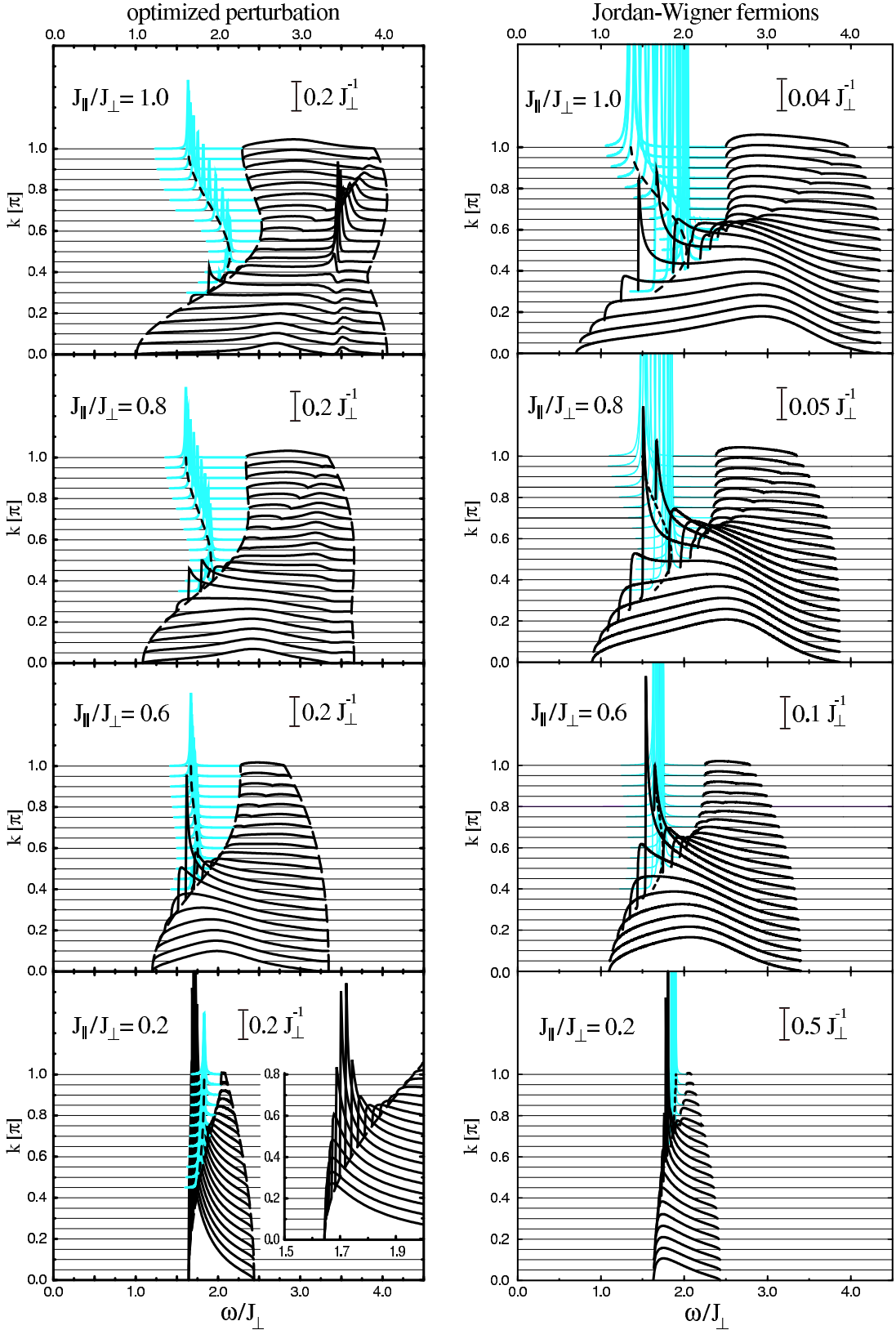,width=14.5cm,clip=}}
  \caption{Momentum dependence of the two-triplet spectral densities with
  $S_{tot}=0$ for $E\parallel c$. Calculations were performed with optimized
  perturbation (left) and with Jordan-Wigner fermions (right) for
  $0 \leq J_\parallel / J_\perp \leq 1$.
  The $k$-resolved and $\omega$-integrated weights for the two theories agree
  within 25\% (note the different scalings of the plots).
  Gray curves represent the bound state, which was divided by 16 and artificially
  broadened by $J_{\perp} / 100$ in the left panels.
  Black curves indicate the continuum.
  Inset of bottom left panel: enlarged view of the continuum, the bound state is
  not shown here.
}
\label{alle}
\end{figure}

In order to determine the optical conductivity one needs to calculate the
momentum-resolved two-particle spectral densities with $S_{\rm tot}=0$
\cite{knetterPRL}. The evolution of the spectral densities from
$J_\parallel /J_\perp = 0.2$ to $J_\parallel /J_\perp = 1$ is plotted for
both theories in Fig.\ \ref{alle}. The optical conductivity as given in
Fig.\ \ref{pint} is obtained by integrating these $k$-resolved curves with
a weight factor $\omega \cdot \sin^4(k/2)$ \cite{windt}. For comparison with
experiment one has to add the phonon energy as a constant shift of the energy
scale (before multiplication by $\omega$).

For $J_\parallel /J_\perp = 0.2$ the elementary triplet still shows only
little dispersion (see Fig.\ \ref{disp}). Therefore the two-particle continuum
is rather narrow; the spectral weight is piled up at the bottom of the
continuum for small momenta $k$, and for large $k$ a bound state is formed below
the continuum (see Fig.\ \ref{alle}). For small ratios of $J_\parallel /J_\perp$
this bound state reaches its maximum energy at the Brillouin zone boundary and
dominates in $\sigma(\omega)$ as a single sharp peak with only a small continuum
contribution at higher energies \cite{jurecka} (see left panels of Fig.\ \ref{pint}).
With increasing $J_\parallel /J_\perp$ the bound state acquires a strong dispersion,
and for $J_\parallel /J_\perp \! \gtrsim \! 0.5$ it shows a maximum at $k\approx \pi/2$
and a minimum at the Brillouin zone boundary (see Fig.\ \ref{alle}). Both give rise
to van Hove singularities in the density of states which cause peaks in $\sigma(\omega)$.
Therefore, the dominant peak observed for $J_\parallel /J_\perp \! = \! 0.2$ splits into
two with increasing $J_\parallel$ (see left panels of Fig.\ \ref{pint}). Comparison of
these spectra with the experimental data offers a very accurate tool to determine the
exchange coupling constants, since the frequencies of the two
peaks depend strongly on the coupling ratio $J_\parallel / J_\perp$ \cite{windt}.
For La$_{5.2}$Ca$_{8.8}$Cu$_{24}$O$_{41}$ we obtain $J_\parallel /J_\perp \! = \! 1.0$
and $J_\parallel \! = \! 1100$~cm$^{-1}$ from the Jordan-Wigner fermions, whereas
perturbation theory yields $J_\parallel /J_\perp \! = \! 1.15$ and
$J_\parallel \! = \! 1080$~cm$^{-1}$. This 15\% discrepancy reflects the differences
between the two theories that were already present in the dispersion of the elementary
triplet (see Fig.\ \ref{disp}). Both theories reproduce the experimental line shape
rather well for both polarization directions. Note that the lower peak is suppressed
for $E\parallel a$ due to symmetry \cite{windt}.

A ratio of $J_\parallel / J_\perp \approx 1$ seems to be in
conflict with several former results of other techniques,
proposing $J_\parallel / J_\perp \gtrsim 1.5$ (see discussion in
\cite{johnston}). Such a rather small value of $J_\perp$ is
suggested by the small spin gap observed
in e.g.\ neutron
scattering \cite{matsuda}. On the basis of our results we can
exclude $J_\parallel /J_\perp > 1.2$ \cite{windt}. Recently, it
was pointed out that the neutron data (i.e., the small spin gap)
are also consistent with an isotropic exchange
$J_\parallel/J_\perp \approx 1 - 1.1$ and $J_\parallel \approx$
900 cm$^{-1}$, if a ring exchange of $J_{\rm cyc} \approx $ 0.15
$J_\parallel$ is taken into account \cite{matsuda}. Adding a
finite ring exchange term reduces the gap at $k=\pi$ and washes
out the dip in the triplet dispersion at small $k$ \cite{matsuda,schmidt}.
Furthermore, the cyclic exchange weakens
the attractive interaction between two rung triplets. We expect
that these two changes, reduction of the dip and of the
attractive interaction, render the theoretical predictions closer
to the experimental findings of optical spectroscopy
\cite{schmidt,knetter2}. Since a $\approx$~15\% ring exchange term is
estimated \cite{matsuda,schmidt} we expect a similar change in
$J_\parallel/J_\perp$.

Finally, we address the evolution of the continuum with
increasing $J_\parallel /J_\perp$ (see Fig.\ \ref{alle}). At
small $k$, the spectral density is broadened strongly by the
increase of the continuum width. Therefore, the $k$=0 Raman
response shows a sharp peak for small $J_\parallel /J_\perp$ and
a broad band for $J_\parallel /J_\perp \! \approx \! 1$ (see Ref.\
\cite{schmidt} for more details). For large momenta one can
observe the opposite, the features within the continuum become
stronger and more pronounced with increasing $J_\parallel
/J_\perp$. The mid-band line of square root singularities which runs from
$k\approx \pi/2$ to $k=\pi$ denotes the upper edge of the
precursor of the spinon continuum that is well known from the
chain limit \cite{knetterPRL}. For $J_\parallel /J_\perp$=1 the
perturbation result (top left panel of Fig.\ \ref{alle}) shows
additional pronounced features within the high energy part of the
continuum \cite{knetterPRL}. The appearance of these features is
related to the existence of the dip in the dispersion of the
elementary triplet at small $k$ (Fig.\ \ref{disp}). Precursors of
these features are present in perturbation theory for
$J_\parallel /J_\perp$=0.8, where the dip in the
dispersion is only small. The Jordan-Wigner fermions do not show
these features.

In the weighted superpositions (Fig.\ \ref{pint}) the perturbative
results for $J_\parallel /J_\perp \ge 0.8$ display a second peak
above $\omega\approx 3J_\perp$ in the high energy continua. The
experimental data (Fig.\ \ref{exptheo}) also display two
peaks in the continuum, independent of the precise background correction.
Yet the experimental peaks lie at higher energies and they are
further apart from each other. On inclusion  of ring exchange the
calculated peaks will be shifted to higher energies since the
attractive interaction is weakened. So we tentatively identify
the two experimental with the two theoretical peaks even though
their shape and mutual distance are not in perfect agreement.
Further investigations including ring exchange are indispensable.

Summarizing, we reported in detail on one of the first observations
of bound states in gapped spin systems. By phonon-assisted
two-triplet absorption we were able to detect a bound
state in the $S=0$ channel resulting from two elementary triplets
in the undoped spin ladder La$_x$Ca$_{14-x}$Cu$_{24}$O$_{41}$.
The bound state exists for momenta $k\gtrapprox0.3\pi$.
Thus, it was decisive that phonons are involved in the optical
excitation process in order to provide the necessary momentum.
Two theoretical methods were employed to trace the evolution of
the spectral densities as function of the coupling ratio
$J_\parallel/J_\perp$. In both approaches the overall position
and weight of the bound state agree well with the measurements. In
experiment and in the perturbative approach two peaks in the
continua are found. Some quantitative differences indicate that
an extension of a simple spin ladder Hamiltonian is necessary. We
argued that a 10-15\% ring exchange term is the appropriate
extension.

This project is supported by the DFG (FR 754/2-1, SP 1073 and SFB
484), by the BMBF (13N6918/1) and by the DAAD within the scope of
PROCOPE.

\onecolumn
\twocolumn

\end{document}